\documentclass{cjaa}                   % preprint, the final version for publication
\usepackage{graphicx}                   %for PS/EPS graphics inclusion, new
\input{epsf.sty}                        %for PS/EPS graphics inclusion, old
\input{psfig.sty}                       %for PS/EPS graphics inclusion, old

\begin{document}

   \title{Fitting Formulae for the Effects of Binary Interactions on Lick Indices and Colours of Stellar Populations
%\footnotetext{$*$ Supported by the National Natural Science Foundation of China.}
}
%   \subtitle{I. Place Your Subtitle Here}

   \volnopage{Vol.0 (200x) No.0, 000--000}      %%preserved for Editor. DOn't remove!
   \setcounter{page}{1}          %%starting page, preserved for Editor. DOn't remove!

   \author{Zhongmu Li
      \inst{1,2}\mailto{}
%% Please move "\mailto{}" to the corresponding author of the paper.
   \and Zhanwen Han
      \inst{1}
      }
   \offprints{Z.-M. Li}                   %% is disabled in fact

   \institute{National Astronomical Observatories/Yunnan
Observatory, the Chinese Academy of Sciences, Kunming, 650011,
China\\
             \email{zhongmu.li@gmail.com}
%% offprint requests will be sent. Note to pair \mailto{} with \email{}
        \and
             Graduate University of the Chinese Academy of
Sciences\\
          }

   \date{Received~~2007 Dec. 17; accepted~~2001~~month day}

   \abstract{
    More than about 50\% stars of galaxies are in binaries, but
    most stellar population studies take single star-stellar
    population (ss-SSP) models, which do not take binary interactions into
    account. In fact, the integrated peculiarities of ss-SSPs are various from
    those of stellar populations with binary interactions (bs-SSPs).
    Therefore, it is necessary to investigate the effects of binary
    interactions on the Lick indices and colours of populations detailedly.
    We show some formulae for calculating the difference between
    the Lick indices and colours of bs-SSPs, and those of ss-SSPs. Twenty-five Lick
    indices and 12 colours are studied in the work.
    The results can be conveniently used for estimating the effects
    of binary interactions on stellar population studies and for
    adding the effects of binary interactions into existing ss-SSP
    models. The results and a few procedures can be obtained on request to the authors or
    via http://www.ynao.ac.cn/$^\sim$bps/zhongmu/download.htm.
   }

   \authorrunning{Zhongmu Li \& Zhanwen Han}            %author_head in even pages
   \titlerunning{Fitting Formulae for Binary Interactions}  % title_head in odd pages

   \maketitle
%% The author head (on even pages) and the title head (on odd pages) will be
%% automatically extracted from \author{} and \title{}. Whenever the title is too long,
%% you will be asked to supply a shorter one by inserting either \authorrunning{} or
%% \titlerunning{} before \maketitle. Anyway, you can specify your own heads in advance.
%%
%%
%% Note: In the following text body of your manuscript, please note several differences from
%%       other major journals:
%% (1) \subsection{Please Capitalize the First Letter of Each Notional Word in Subsection Title}
%% (2) Please Capitalize the First Letter of Each Notional Word in table's caption

%
%________________________________________________ sections below
%
\section{Introduction}           %% first-level sections will be auto-capitalized
\label{sect:intro}
%\hspace{15pt}%                   %% preserved for Editor
In the golden era for studying the formation and evolution of
galaxies, evolutionary stellar population synthesis has been an
important technique for such works, as some stellar peculiarities
(e.g., stellar age and metallicity) of galaxies can be determined
via this technique. Many stellar population synthesis models, e.g.,
\cite{Worthey:1994}, \cite{Buzzoni:1995}, \cite{Bressan:1994},
\cite{Vazquez:2005}, \cite{Bruzual:2003}, \cite{Fioc:1997},
\cite{Vazdekis:2003}, \cite{Delgado:2005}, and \cite{Zhang:2005hr},
were brought forward and have been widely used for stellar
population studies. However, the above models except the one of
\cite{Zhang:2005hr} are single star-stellar population (ss-SSP)
models that did not take the effects of binary interactions into
account. This is different from the real populations of galaxies and
star clusters. According to the results of \cite{Han:2001}, more
than 50\% stars of the Galaxy are in binaries and evolve differently
from single stars. The real stellar populations of galaxies and star
clusters consist of not only single stars, but also binary stars.
Binary evolution can affect the results of stellar population
synthesis significantly, especially those relating to UV bands, see,
e.g., \cite{Han:2007}. Therefore, the effects of binary evolution
should be taken into account when modeling the stellar populations
of galaxies and star clusters.

A few works have been tried to give some investigations about the
effects of binary evolution on stellar population synthesis. For
example, \cite{Zhang:2005hr} tried to model populations via binary
stars. In addition, \cite{Li:2007database} built an isochrone
database for quickly modeling binary star-stellar populations
(bs-SSPs) and a rapid model (hereafter $RPS$ model) for both ss-SSPs
and bs-SSPs. In special, \cite{Li:2007binaryeffects} investigated
the detailed effects of binary interactions on the results of
stellar population synthesis and the results of stellar population
studies. The results can help us to understand how the results
obtained via ss-SSPs are different from those obtained via bs-SSPs,
when taking the H$\beta$--[MgFe] (\cite{Thomas:2003}) and two-colour
methods. According to the results of \cite{Li:2007binaryeffects},
when we use ss-SSP models to measure the stellar ages and
metallicities of galaxies, we will obtain obviously less ages or
less metallicities compared to the real values of populations, using
H$\beta$--[MgFe] and two-colour methods, respectively. However,
there is no clear relation between the real metallicities and fitted
(via ss-SSPs) results of populations. One please refer to
\cite{Li:2007binaryeffects} for more details. In this case, it is
difficult to get more accurate information about the stellar
metallicities of galaxies via ss-SSP models, and then the chemical
evolution of galaxies. Furthermore, the previous work only shows the
results for H$\beta$--[MgFe] method, when taking Lick indices for
works, but some other methods and indices are also used in
investigations. Thus it is necessary to investigate the effects of
binary interactions on the results of stellar population studies
obtained via various Lick indices further. The metallicity range of
above bs-SSP models (\cite{Zhang:2005hr}, \cite{Li:2007database})
seems not wide enough (see \cite{Li:2006}), as it only covers the
metallicity range poorer than 0.03 ($Z \le$ 0.03). If we can give
the relation between the effects of binary interactions and the
stellar-population parameters (age and metallicity), we will be able
to understand the populations of galaxies and star clusters further,
and more detailed investigations about galaxy formation and
evolution will have in the future. Therefore, it is valuable to
study how the effects of binary interactions on integrated
peculiarities of populations change with stellar age and
metallicity. We have a try in this work. As a result, a few formulae
for describing the relations between the effects of binary
interactions on 25 Lick indices and 12 colours, and the ages and
metallicities of populations are presented.

The structure of the paper is as follows. In Sect. 2 we introduce
the ss-SSP and bs-SSP models used in the paper. In Sect. 3 we show
the fitting formulae for the changes of 25 Lick indices caused by
binary interactions, comparing to those of ss-SSPs. In Sect. 4 we
give similar investigations to 12 colours of populations. Finally,
we give our discussion and conclusion in Sect. 5.

%% ChJAA editors DID NOT use \cite{} for citation, \ref and \label for
%% cross-references of Table/Figure in publication version.
%% ChJAA editors prefered you giving a citation as 'Michel et al. 1992', and
%% writting Table~1 or Fig.~1 and so forth. However, that will make authors
%% inconvenient in adjusting/adding/removing text, tables or figures. Anyway,
%% authors can use \cite, \citep and \citet as widely used in other journals.
%% ChJAA editors are moving to use a more flexible LaTeX source.

\section{Stellar population synthesis model used in the paper}
\label{sect:Obs}
%\hspace{15pt}%                   %% preserved for Editor
The $RPS$ model of \cite{Li:2007database} is used in the
investigation, because there is no more suitable model. The model
calculated the integrated peculiarities (0.3 $\rm \AA$ SEDs, Lick
indices and colours) of both bs-SSPs and ss-SSPs with two widely
used initial mass functions (IMFs) (Salpeter and Chabrier IMFs).
Each bs-SSP contains about 50\% stars that are in binaries with
orbital periods less than 100\,yr (the typical value of the Galaxy,
see \cite{Han:1995}). Binary interactions such as mass transfer,
mass accretion, common-envelope evolution, collisions, supernova
kicks, angular momentum loss mechanism, and tidal interactions are
considered when evolving binaries via the rapid stellar evolution
code of \cite{Hurley:2002}. Therefore, the $RPS$ model is suitable
for studying the effects of binary interactions on stellar
population synthesis. The details about the model can be seen in the
paper of \cite{Li:2007database}. For convenience, we take stellar
populations with Salpeter IMF for our standard investigations in the
work, but the results obtained via populations with Chabrier IMF are
also presented.

\section{Fitting formulae for effects of binary interactions on Lick indices}
\label{sect:data}
%\hspace{15pt}%                   %% preserved for Editor
Lick indices are the most widely used indices in stellar population
studies, because they can disentangle the well-known
age--metallicity degeneracy (\cite{Worthey:1994}). Making use of an
age-sensitive index (e.g., H$\beta$) and a metallicity-sensitive
index (e.g., [MgFe], see \cite{Thomas:2003}), the stellar age and
metallicity of a population can be determined. Thus to investigate
the effects of binary interactions on the Lick indices of stellar
populations is important. The work of \cite{Li:2007binaryeffects}
showed that binary interactions make the H$\beta$ index less while
some metal-line indices larger compared to those of ss-SSPs. It
leads to less age estimate when we take ss-SSPs for works. However,
in that work, only the results obtained via H$\beta$--[MgFe] method
are compared to the real values of populations. Some other Lick
indices, e.g., Mg2, H$\delta_{\rm A}$, and H$\gamma_{\rm A}$, are
also used in studies (e.g., \cite{Gallazzi:2005}). Therefore, it is
necessary to study the effects of binary interactions on more Lick
indices and give the quantitative relations between binary effects
and stellar-population parameters. Here we study on 25 widely used
indices and fit the relations between the changes caused by binary
interactions and the stellar-population parameters (age and
metallicity), via a polynomial fitting method. The results can be
used to calculate the differences between 25 Lick indices, and the
errors are small (typically less than 0.03 ${\rm \AA}$ or mag). All
Lick indices are on the Lick system (see, e.g.,
\cite{Worthey:1994licksystem}). The changes of Lick indices caused
by binary interactions can be calculated from stellar age and
metallicity, by
\begin{equation}
    \Delta I = \sum_{i=1}^{5}
({\rm C}_{i{\rm 1}} + {\rm C}_{i{\rm 2}}Z + {\rm C}_{i{\rm 3}}Z^{\rm
2})t^{i-1},
\end{equation}
where $\Delta I$ is the change of a Lick index caused by binary
interactions, and $Z$ is stellar metallicity while $t$ is stellar
age. The detailed coefficients for our standard investigation are
shown in Tables 1 and 2. Those for populations with Chabrier IMF are
shown in the Appendix. For clearly, in Figs. 1, 2, and 3, we compare
the changes calculated by equation (1) with the original values
obtained in the work. Note that we only show the fittings for 12
widely used Lick indices here, because the fittings for other
indices are similar. As we see, for the indices shown, the values
calculated by the above equation are consistent with those obtained
directly by comparing the Lick indices of bs-SSPs and ss-SSPs, with
typical errors of 0.03\,${\rm \AA}$ or mag. Therefore, the fitting
formulae presented can be used to calculate the differences of Lick
indices of bs-SSPs and ss-SSPs, using the age and metallicity of
populations. In addition, the results show that as what were shown
in the paper of \cite{Li:2007binaryeffects}, binary interactions
make age-sensitive indices (e.g., H$\beta$, $H\delta_{\rm A}$,
$H\delta_{\rm F}$, $H\gamma_{\rm A}$, $H\gamma_{\rm F}$) of a bs-SSP
larger than that of an ss-SSP, which has the same age and
metallicity as the bs-SSP, while the interactions make
metallicity-sensitive indices (e.g., Mg or Fe indices) of a bs-SSP
less than that of its corresponding (with the same age and
metallicity) ss-SSP. The differences between Lick indices of bs-SSPs
and ss-SSPs increase with age when stellar age is small ($<$ about
2.5\,Gyr), and they decrease with age for stellar ages larger than
about 2.5\,Gyr. As a whole, the values calculated via the fitting
formulae obtained by the paper reproduce the evolution of the
difference between Lick indices of bs-SSPs and ss-SSPs.
\begin{table}[h]
\caption[]{Coefficients for equation (1). The coefficients are
obtained via stellar populations with Salpeter IMF and can be used
for populations younger than 4\,Gyr (Age $<$ 4\,Gyr).} \label{Tab:4}
\begin{center}\begin{tabular}{lcrrrrr}
\hline\hline\noalign{\smallskip}%\scriptsize
Index&$j$ &${\rm C_{1j}}$ &${\rm C_{2j}}$ &${\rm C_{3j}}$ &${\rm C_{4j}}$ &${\rm C_{5j}}$\\
\hline
                  &1 &    0.0061004 &   -0.0124136 &    0.0044131 &   -0.0010174 &    0.0000862 \\
CN$_{\rm 1}$      &2 &   -0.7411409 &    3.6764976 &   -3.4540761 &    0.9481077 &   -0.0789850 \\
                  &3 &   30.8122734 & -119.0088796 &   98.2842545 &  -25.5632396 &    2.0732871 \\
\hline                                                                                          \\
                  &1 &    0.0031344 &   -0.0060648 &    0.0018661 &   -0.0005851 &    0.0000607 \\
CN$_{\rm 2}$      &2 &   -0.3472402 &    2.6820361 &   -2.8891002 &    0.8270163 &   -0.0703722 \\
                  &3 &   20.0763558 &  -92.9530724 &   84.0072412 &  -22.6637506 &    1.8767287 \\
\hline                                                                                          \\
                  &1 &    0.0116927 &   -0.0652076 &    0.0542984 &   -0.0142708 &    0.0011505 \\
Ca4227            &2 &    2.6086437 &    1.4449921 &   -5.0762513 &    1.6986709 &   -0.1529064 \\
                  &3 &  -71.6010754 &   65.3934655 &   -2.6926390 &   -6.3483260 &    0.8599647 \\
\hline                                                                                          \\
                  &1 &    0.5779073 &   -1.3565672 &    0.6261188 &   -0.1240817 &    0.0083961 \\
G4300             &2 &  -50.8998137 &  134.5958812 &  -94.7989188 &   23.2438808 &   -1.8011864 \\
                  &3 & 1740.9893518 &-4249.3881941 & 2828.2830757 & -668.7058103 &   50.9809128 \\
\hline                                                                                          \\
                  &1 &    0.1013725 &   -0.1526486 &   -0.0164836 &    0.0124241 &   -0.0013244 \\
Fe4383            &2 &   11.0894652 &   -9.0004221 &  -12.0126956 &    4.9108205 &   -0.4569935 \\
                  &3 &  -75.5186140 & -252.7173536 &  549.1707675 & -171.1008580 &   14.7310943 \\
\hline                                                                                          \\
                  &1 &   -0.0047432 &    0.0186972 &   -0.0197387 &    0.0040785 &   -0.0002709 \\
Ca4455            &2 &    6.5346797 &  -11.0259844 &    2.2737411 &    0.1733499 &   -0.0467838 \\
                  &3 &  -98.3181765 &   98.4157661 &   49.1921063 &  -27.0189711 &    2.8005540 \\
\hline                                                                                          \\
                  &1 &    0.0550772 &   -0.1274444 &    0.0222163 &   -0.0009561 &   -0.0000671 \\
Fe4531            &2 &   18.2809537 &  -32.2387726 &   11.1472063 &   -1.0492667 &   -0.0001559 \\
                  &3 & -313.2684866 &  374.3927838 &  -24.2302367 &  -31.1448033 &    4.4257825 \\
\hline                                                                                          \\
                  &1 &   -0.0717857 &    0.2203558 &   -0.1810129 &    0.0435325 &   -0.0033410 \\
Fe4668            &2 &   27.2930010 &  -66.0581381 &   35.0658395 &   -6.9644712 &    0.4740158 \\
                  &3 & -444.6620029 &  913.3400658 & -347.7358123 &   51.9244201 &   -2.7150328 \\
\hline                                                                                          \\
                  &1 &   -0.4120752 &    0.9250669 &   -0.3741119 &    0.0622541 &   -0.0036591 \\
H$_\beta$         &2 &   26.4414835 &  -74.8894421 &   48.1881455 &  -11.2855830 &    0.8626001 \\
                  &3 & -805.0295817 & 2280.7328848 &-1494.4764420 &  349.3195439 &  -26.6678020 \\
\hline                                                                                          \\
                  &1 &   -0.0313311 &   -0.0500534 &   -0.0035580 &   -0.0037280 &    0.0007775 \\
Fe5015            &2 &   61.9785871 & -107.1406403 &   40.6549692 &   -4.5784440 &    0.0744308 \\
                  &3 &-1490.9225648 & 2286.4644788 & -734.9273433 &   46.6787692 &    3.5030950 \\
\hline                                                                                          \\
                  &1 &   -0.0011669 &    0.0020035 &   -0.0025450 &    0.0006800 &   -0.0000553 \\
Mg$_{\rm 1}$      &2 &    1.0977834 &   -1.6954874 &    0.4725610 &   -0.0265407 &   -0.0021062 \\
                  &3 &  -23.8411177 &   30.5518215 &   -4.5747311 &   -1.0811535 &    0.1935160 \\
\hline                                                                                          \\
                  &1 &   -0.0036972 &    0.0016391 &   -0.0002221 &   -0.0002245 &    0.0000340 \\
Mg$_{\rm 2}$      &2 &    2.6546194 &   -4.0910074 &    1.1294327 &   -0.0429887 &   -0.0081361 \\
                  &3 &  -63.2072524 &   93.1999501 &  -25.0701341 &    0.5006265 &    0.2481080 \\
\hline                                                                                          \\
                  &1 &   -0.1159519 &    0.0564072 &    0.0367218 &   -0.0207147 &    0.0023015 \\
Mg$_{\rm b}$      &2 &   36.6456419 &  -47.5077586 &    8.8070472 &    1.2388375 &   -0.2750762 \\
                  &3 & -857.5383388 & 1080.0808279 & -207.5498638 &  -26.8404813 &    6.2521890 \\
\hline                                                                                          \\
                  &1 &    0.1443967 &   -0.2658498 &    0.0907173 &   -0.0111382 &    0.0003331 \\
Fe5270            &2 &   -8.5450001 &   12.6759249 &   -7.5470939 &    1.5641500 &   -0.0991560 \\
                  &3 &  306.8420600 & -637.6523032 &  403.3286320 &  -89.3943857 &    6.3626512 \\
\noalign{\smallskip}

\noalign{\smallskip}\hline
\end{tabular}\end{center}
\end{table}

\addtocounter{table}{-1}
\begin{table}[b]
\centering \caption[]{--continued.} \label{Tab:4}
\begin{center}\begin{tabular}{lcrrrrr}
\hline\hline\noalign{\smallskip}%\scriptsize
Index&$j$ &${\rm C_{1j}}$ &${\rm C_{2j}}$ &${\rm C_{3j}}$ &${\rm C_{4j}}$ &${\rm C_{5j}}$\\
\hline                                                                                        \\
                  &1 &    0.1121199 &   -0.2143513 &    0.0776458 &   -0.0095392 &    0.0002385 \\
Fe5335            &2 &   -6.1989089 &   11.7530508 &   -8.2144476 &    1.6722406 &   -0.0990246 \\
                  &3 &  144.7440581 & -352.4412737 &  218.9898664 &  -44.2594619 &    2.7694391 \\
\hline                                                                                          \\
                  &1 &   -0.0167597 &   -0.0094937 &   -0.0069105 &    0.0022746 &   -0.0001900 \\
Fe5406            &2 &   18.1518764 &  -28.2532163 &    9.0539150 &   -0.8243859 &   -0.0014094 \\
                  &3 & -423.9701730 &  596.6530611 & -158.4751506 &    4.4556600 &    1.3344343 \\
\hline                                                                                          \\
                  &1 &    0.0643501 &   -0.1189715 &    0.0352640 &   -0.0029869 &   -0.0000373 \\
Fe5709            &2 &   -6.8927462 &    9.4915829 &   -2.5896330 &    0.0686243 &    0.0256903 \\
                  &3 &  206.1312747 & -349.9572257 &  151.0604011 &  -22.2154935 &    0.9194484 \\
\hline                                                                                          \\
                  &1 &   -0.0193549 &    0.0201941 &   -0.0227794 &    0.0068166 &   -0.0006059 \\
Fe5782            &2 &    5.8734339 &   -8.3105593 &    3.3800221 &   -0.6333537 &    0.0435585 \\
                  &3 & -145.6309054 &  234.0700497 & -130.6085667 &   28.1896236 &   -2.0267932 \\
\hline                                                                                          \\
                  &1 &   -0.1041283 &    0.0782159 &   -0.0131910 &   -0.0028878 &    0.0005590 \\
Na$_{\rm D}$      &2 &   35.8174765 &  -52.1821670 &   16.3021886 &   -1.2441912 &   -0.0404260 \\
                  &3 & -928.3347064 & 1387.5693179 & -484.7687597 &   49.3839423 &   -0.3018862 \\
\hline                                                                                          \\
                  &1 &   -0.0046972 &   -0.0031675 &    0.0086166 &   -0.0031273 &    0.0003031 \\
TiO$_{\rm 1}$     &2 &    2.5376502 &   -3.1526728 &    0.3308384 &    0.2054016 &   -0.0316500 \\
                  &3 &  -70.3829758 &   95.4015845 &  -17.9727317 &   -3.0711691 &    0.6438944 \\
\hline                                                                                          \\
                  &1 &   -0.0089825 &   -0.0020920 &    0.0096106 &   -0.0036755 &    0.0003656 \\
TiO$_{\rm 2}$     &2 &    3.9165850 &   -4.9045789 &    0.8074595 &    0.2009710 &   -0.0375203 \\
                  &3 & -108.7948038 &  146.4687943 &  -32.9218404 &   -2.5469339 &    0.7759394 \\
\hline                                                                                          \\
                  &1 &   -0.3400174 &    0.7802154 &   -0.3437893 &    0.0824056 &   -0.0066930 \\
H$\delta_{\rm A}$ &2 &   36.1459627 & -178.0526116 &  166.3449941 &  -45.4458738 &    3.7573350 \\
                  &3 &-1381.3144000 & 5283.2867289 &-4334.0624242 & 1125.3443000 &  -90.9465139 \\
\hline                                                                                          \\
                  &1 &   -0.7586478 &    1.6470466 &   -0.6800223 &    0.1306246 &   -0.0086650 \\
H$\gamma_{\rm A}$ &2 &   58.8625436 & -210.0639058 &  174.8989489 &  -45.3356756 &    3.6089852 \\
                  &3 &-2043.5700337 & 6433.0522877 &-4883.0699797 & 1207.4959655 &  -93.9809592 \\
\hline                                                                                          \\
                  &1 &   -0.3060721 &    0.7125074 &   -0.3205038 &    0.0683519 &   -0.0050475 \\
H$\delta_{\rm F}$ &2 &   30.6677991 & -121.8747783 &  101.4148311 &  -26.5930774 &    2.1537733 \\
                  &3 &-1053.7758076 & 3611.9972771 &-2723.9526032 &  682.2344274 &  -54.0667213 \\
\hline                                                                                          \\
                  &1 &   -0.4440136 &    0.9835898 &   -0.4158944 &    0.0801404 &   -0.0053903 \\
H$\gamma_{\rm F}$ &2 &   35.2952498 & -120.5372324 &   95.1523223 &  -24.3827679 &    1.9426707 \\
                  &3 &-1145.3060328 & 3586.2350776 &-2658.3958905 &  655.6380015 &  -51.2487311 \\
\noalign{\smallskip}

\noalign{\smallskip}\hline
\end{tabular}\end{center}
\end{table}

\begin{table}[b]
\caption[]{Similar to Table 1, but for stellar populations older
than 4\,Gyr (Age $\geq$ 4\,Gyr).} \label{Tab:4}
\begin{center}\begin{tabular}{lcrrrrr}
\hline\hline\noalign{\smallskip}%\scriptsize
Index&$j$ &${\rm C_{1j}}$ &${\rm C_{2j}}$ &${\rm C_{3j}}$ &${\rm C_{4j}}$ &${\rm C_{5j}}$\\
\hline
                  &1 &    0.0040188 &   -0.0087338 &    0.0012365 &   -0.0000628 &    0.0000011 \\
CN$_{\rm 1}$      &2 &    0.0699149 &   -0.5550663 &    0.1605598 &   -0.0163788 &    0.0005280 \\
                  &3 &   -4.2835098 &   11.3189732 &   -2.8137761 &    0.2856955 &   -0.0093980 \\
\hline                                                                                          \\
                  &1 &    0.0031607 &   -0.0068581 &    0.0009171 &   -0.0000386 &    0.0000004 \\
CN$_{\rm 2}$      &2 &    0.1358253 &   -0.6142418 &    0.1616565 &   -0.0161493 &    0.0005224 \\
                  &3 &   -5.8469187 &   12.5102930 &   -2.9058639 &    0.2889257 &   -0.0095349 \\
\hline                                                                                          \\
                  &1 &    0.0024019 &   -0.0096646 &    0.0028396 &   -0.0003487 &    0.0000134 \\
Ca4227            &2 &    0.1289259 &   -0.3224843 &   -0.1484763 &    0.0361270 &   -0.0016691 \\
                  &3 &   30.5865753 &  -70.1598679 &   19.3765757 &   -2.1609623 &    0.0788180 \\
\hline                                                                                          \\
                  &1 &    0.1365771 &   -0.4119753 &    0.0710301 &   -0.0055584 &    0.0001697 \\
G4300             &2 &   -3.6369924 &    2.4500207 &    0.5456707 &   -0.0466573 &   -0.0003761 \\
                  &3 &   28.2611775 &  -85.3355224 &   18.7904056 &   -2.1578069 &    0.1062849 \\
\hline                                                                                          \\
                  &1 &    0.0695695 &   -0.1937149 &    0.0295201 &   -0.0017795 &    0.0000416 \\
Fe4383            &2 &    1.5938861 &   -9.3601744 &    2.3994053 &   -0.2263281 &    0.0068871 \\
                  &3 &  -79.2726970 &  189.8780408 &  -38.5276774 &    3.5614204 &   -0.1108842 \\
\hline                                                                                          \\
                  &1 &    0.0113051 &   -0.0260901 &    0.0031068 &   -0.0001249 &    0.0000015 \\
Ca4455            &2 &    0.1825335 &   -2.1257283 &    0.6162648 &   -0.0626674 &    0.0020383 \\
                  &3 &  -18.3660159 &   52.8898164 &  -13.1167084 &    1.3350018 &   -0.0447127 \\
\hline                                                                                          \\
                  &1 &    0.0241263 &   -0.1055130 &    0.0195680 &   -0.0014870 &    0.0000418 \\
Fe4531            &2 &   -0.8767742 &   -0.8527878 &    0.3649493 &   -0.0354193 &    0.0009578 \\
                  &3 &   25.2479921 &  -36.0000062 &    6.6317767 &   -0.5259928 &    0.0185966 \\
\hline                                                                                          \\
                  &1 &    0.0179749 &   -0.0676067 &    0.0083280 &   -0.0002240 &   -0.0000019 \\
Fe4668            &2 &    0.2262582 &   -4.4121200 &    1.5353953 &   -0.1944241 &    0.0071387 \\
                  &3 &  -79.5135908 &  227.1468134 &  -55.0061793 &    6.0957633 &   -0.2199873 \\
\hline                                                                                          \\
                  &1 &   -0.0625565 &    0.2708823 &   -0.0566269 &    0.0045862 &   -0.0001317 \\
H$_\beta$         &2 &    2.8700663 &   -5.7104589 &    0.9274182 &   -0.0705995 &    0.0022889 \\
                  &3 &  -40.1033590 &   83.2554372 &  -18.0609224 &    1.5794001 &   -0.0541752 \\
\hline                                                                                          \\
                  &1 &    0.0326274 &   -0.1721448 &    0.0331789 &   -0.0025354 &    0.0000699 \\
Fe5015            &2 &   -1.1476034 &   -1.5307235 &    0.5946861 &   -0.0674897 &    0.0022440 \\
                  &3 &    7.2743924 &   22.9005565 &   -3.9738686 &    0.4451390 &   -0.0154651 \\
\hline                                                                                          \\
                  &1 &    0.0002583 &   -0.0028680 &    0.0005454 &   -0.0000406 &    0.0000011 \\
Mg$_{\rm 1}$      &2 &    0.0303274 &   -0.1102308 &    0.0115624 &   -0.0003957 &   -0.0000009 \\
                  &3 &   -0.0290541 &    0.0201172 &    0.2573529 &   -0.0302840 &    0.0011283 \\
\hline                                                                                          \\
                  &1 &    0.0003494 &   -0.0047501 &    0.0010250 &   -0.0000857 &    0.0000026 \\
Mg$_{\rm 2}$      &2 &    0.0392009 &   -0.1979893 &    0.0208337 &   -0.0000621 &   -0.0000425 \\
                  &3 &    0.4638829 &   -0.5123346 &    0.6866533 &   -0.0971528 &    0.0039890 \\
\hline                                                                                          \\
                  &1 &   -0.0037660 &   -0.0157034 &    0.0047387 &   -0.0004621 &    0.0000160 \\
Mg$_{\rm b}$      &2 &    0.9698109 &   -4.0019894 &    0.5949929 &   -0.0254183 &    0.0000838 \\
                  &3 &  -10.4139848 &   42.6783675 &   -1.0978610 &   -0.5308182 &    0.0325709 \\
\hline                                                                                          \\
                  &1 &    0.0135064 &   -0.0676152 &    0.0131365 &   -0.0009717 &    0.0000261 \\
Fe5270            &2 &   -0.3787006 &   -1.1509906 &    0.2995794 &   -0.0297612 &    0.0009629 \\
                  &3 &    2.9257542 &   20.0839626 &   -2.9644946 &    0.2873734 &   -0.0103891 \\
\noalign{\smallskip}

\noalign{\smallskip}\hline
\end{tabular}\end{center}
\end{table}

\addtocounter{table}{-1}
\begin{table}[h]
\centering \caption[]{--continued.} \label{Tab:4}
\begin{center}\begin{tabular}{lcrrrrr}
\hline\hline\noalign{\smallskip}%\scriptsize
Index&$j$ &${\rm C_{1j}}$ &${\rm C_{2j}}$ &${\rm C_{3j}}$ &${\rm C_{4j}}$ &${\rm C_{5j}}$\\
\hline                                                                                        \\
                  &1 &    0.0100735 &   -0.0513031 &    0.0112253 &   -0.0009878 &    0.0000303 \\
Fe5335            &2 &   -0.3764122 &   -1.1131799 &   -0.1092792 &    0.0392520 &   -0.0018016 \\
                  &3 &   46.0187216 &  -72.8319926 &   27.5059454 &   -3.2158884 &    0.1133870 \\
\hline                                                                                          \\
                  &1 &    0.0061912 &   -0.0466338 &    0.0093712 &   -0.0007327 &    0.0000206 \\
Fe5406            &2 &    0.0934871 &   -1.3125122 &    0.1939528 &   -0.0108703 &    0.0001877 \\
                  &3 &    1.0004186 &   10.2495446 &    1.2408277 &   -0.2790750 &    0.0120963 \\
\hline                                                                                          \\
                  &1 &    0.0084714 &   -0.0287635 &    0.0050787 &   -0.0003500 &    0.0000089 \\
Fe5709            &2 &   -0.4407574 &    0.0360397 &    0.0939359 &   -0.0153071 &    0.0006047 \\
                  &3 &   -0.8112771 &   19.1722212 &   -4.5075237 &    0.4973342 &   -0.0184540 \\
\hline                                                                                          \\
                  &1 &   -0.0002944 &   -0.0149229 &    0.0032162 &   -0.0003063 &    0.0000099 \\
Fe5782            &2 &    0.2419052 &   -0.2990725 &   -0.1171528 &    0.0246488 &   -0.0010394 \\
                  &3 &   20.8520931 &  -52.8185426 &   16.2838692 &   -1.7500469 &    0.0595642 \\
\hline                                                                                          \\
                  &1 &   -0.0022021 &   -0.0380672 &    0.0095990 &   -0.0008472 &    0.0000260 \\
Na$_{\rm D}$      &2 &    0.3464177 &   -0.9169326 &   -0.0149337 &    0.0195706 &   -0.0010391 \\
                  &3 &   14.6152128 &  -38.0493377 &   12.0162242 &   -1.3802231 &    0.0519050 \\
\hline                                                                                          \\
                  &1 &   -0.0004048 &   -0.0010965 &    0.0002912 &   -0.0000264 &    0.0000008 \\
TiO$_{\rm 1}$     &2 &    0.0098709 &   -0.0060943 &   -0.0015886 &    0.0001050 &   -0.0000058 \\
                  &3 &    0.4022334 &   -0.0305906 &   -0.0854966 &    0.0157629 &   -0.0004537 \\
\hline                                                                                          \\
                  &1 &   -0.0006406 &   -0.0026937 &    0.0006705 &   -0.0000589 &    0.0000018 \\
TiO$_{\rm 2}$     &2 &    0.0411593 &    0.0211796 &   -0.0159178 &    0.0015739 &   -0.0000518 \\
                  &3 &   -0.4009774 &   -0.3431377 &    0.1546906 &   -0.0093757 &    0.0003560 \\
\hline                                                                                          \\
                  &1 &   -0.1822723 &    0.4528323 &   -0.0619613 &    0.0036523 &   -0.0000907 \\
H$\delta_{\rm A}$ &2 &   -9.3242935 &   26.8376168 &   -7.6220614 &    0.6779245 &   -0.0184436 \\
                  &3 &  273.5136304 & -384.4491796 &   87.5549531 &   -6.8624029 &    0.1483407 \\
\hline                                                                                          \\
                  &1 &   -0.2347224 &    0.6562770 &   -0.0985331 &    0.0067082 &   -0.0001900 \\
H$\gamma_{\rm A}$ &2 &   -8.3702686 &   24.2585985 &   -7.8246936 &    0.7094328 &   -0.0186995 \\
                  &3 &  392.7692657 & -531.6987235 &  129.8789712 &  -11.2146737 &    0.2783352 \\
\hline                                                                                          \\
                  &1 &   -0.1106862 &    0.2930026 &   -0.0443820 &    0.0029922 &   -0.0000815 \\
H$\delta_{\rm F}$ &2 &   -3.5060331 &   10.7215430 &   -3.5595302 &    0.3279227 &   -0.0089577 \\
                  &3 &  145.3867861 & -162.5243753 &   44.0186086 &   -3.7498376 &    0.0877458 \\
\hline                                                                                          \\
                  &1 &   -0.1284349 &    0.3643152 &   -0.0583081 &    0.0041664 &   -0.0001196 \\
H$\gamma_{\rm F}$ &2 &   -1.3760493 &    7.0593899 &   -2.8690465 &    0.2678504 &   -0.0067945 \\
                  &3 &  129.2354201 & -186.3215690 &   49.9193880 &   -4.3279933 &    0.0997893 \\
\noalign{\smallskip}

\noalign{\smallskip}\hline
\end{tabular}\end{center}
\end{table}

%
% one-column-wide figure(occupies half-width of a page)
%  -- This is an old way of graphics inclusion with psfig.sty
%------------------------------------------------------------ Fig1: lightcurve
\begin{figure}
   \vspace{2mm}
   \begin{center}
   \hspace{3mm}\psfig{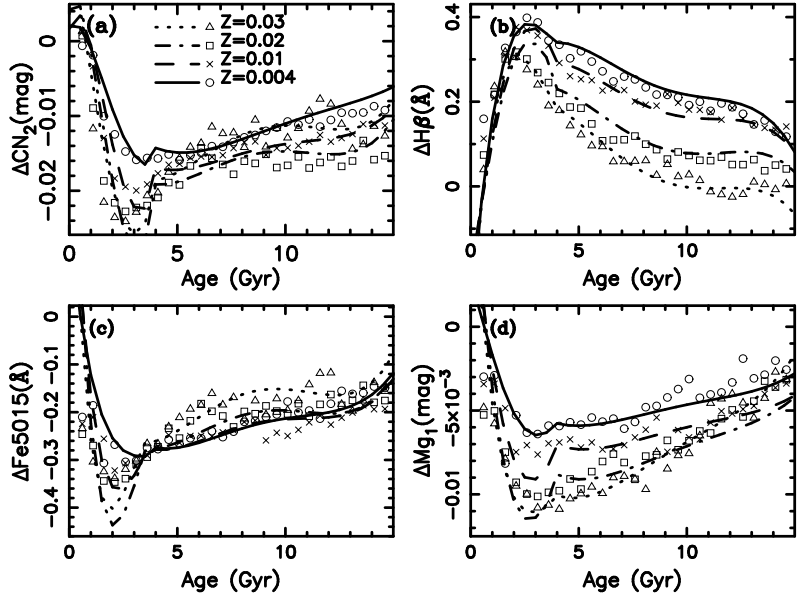}
   \parbox{180mm}{{\vspace{2mm} }}
   \caption{Comparison of fitted and original values for the effects of binary interactions on four Lick indices.
   Circles, crosses, squares, and triangles are for the metallicities of $Z$ = 0.004, 0.01, 0.02, and 0.03, respectively.
   Solid, dashed, dash-dotted, and dotted lines show the fittings for the above metallicities,
   respectively. The values of y-axes are calculated by subtracting
   the Lick indices of a bs-SSP from that of the ss-SSP which has
   the same age and metallicity as the bs-SSP.
   Panels a), b), c), and d) are for CN$_{\rm 2}$, H$\beta$, Fe5015, and Mg$_{\rm 1}$, respectively.}
   \label{Fig:lightcurve-ADAri}
   \end{center}
\end{figure}
%

%
% one-column-wide figure(occupies half-width of a page)
%  -- This is an old way of graphics inclusion with psfig.sty
%------------------------------------------------------------ Fig1: lightcurve
\begin{figure}
   \vspace{2mm}
   \begin{center}
   \hspace{3mm}\psfig{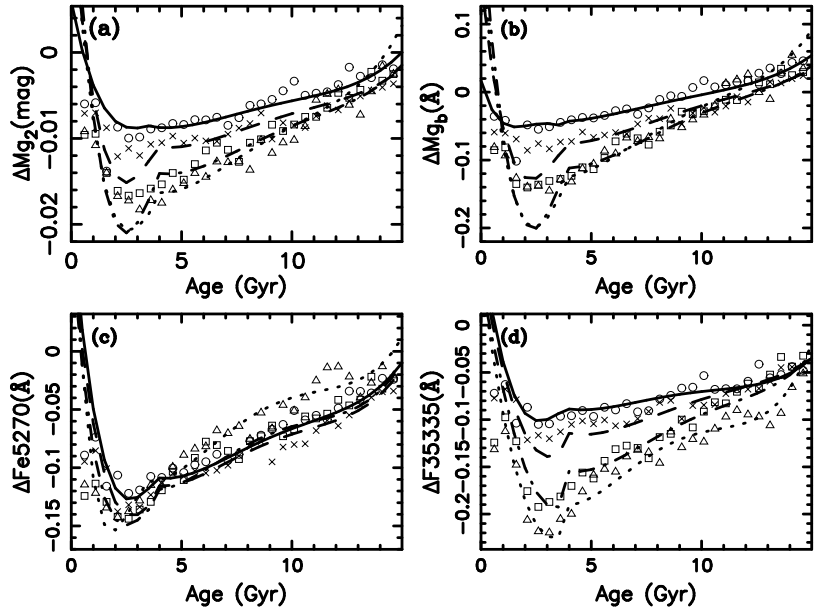}
   \parbox{180mm}{{\vspace{2mm} }}
   \caption{Similar to Fig. 1, but for Mg$_{\rm 2}$, Mg$_{\rm b}$, Fe5270, and Fe5335.}
   \label{Fig:lightcurve-ADAri}
   \end{center}
\end{figure}
%

%
% one-column-wide figure(occupies half-width of a page)
%  -- This is an old way of graphics inclusion with psfig.sty
%------------------------------------------------------------ Fig1: lightcurve
\begin{figure}
   \vspace{2mm}
   \begin{center}
   \hspace{3mm}\psfig{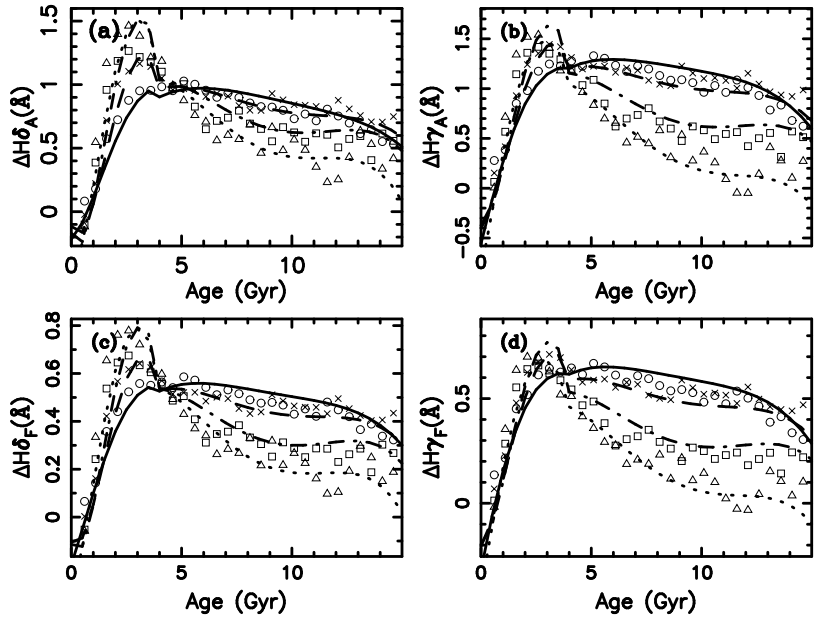}
   \parbox{180mm}{{\vspace{2mm} }}
   \caption{Similar to Fig. 1, but for H$\delta_{\rm A}$, H$\gamma_{\rm A}$, H$\delta_{\rm F}$, and H$\gamma_{\rm F}$.}
   \label{Fig:lightcurve-ADAri}
   \end{center}
\end{figure}

\section{Fitting formulae for effects of binary interactions on colours}
\label{sect:data}
%\hspace{15pt}%                   %% preserved for Editor
Because colours can also be used for stellar population studies, we
fitted the formulae for calculating the changes in colours of
populations that result from binary interactions. One can refer to,
e.g., \cite{Li:2007potentialcolors, Li:2007effects,
Li:2007colourpairs, Li:2007binaryeffects} for the application of
colours in stellar population studies. Colours on Johnson system,
those on the photometry system of Sloan Digital Sky Survey
(hereafter SDSS-$ugriz$ system), and some composite ones that
consist of a Johnson magnitude and an SDSS-$ugriz$ magnitude are
studied. We only study the colours of populations with $Z \geq$
0.004, as it seems difficult to determine the stellar age and
metallicity of metal-poor (e.g., $Z <$ 0.008) populations via
colours under the typical observational uncertainties
(\cite{Li:2007colourpairs}) and metallicity affect the colours of
metal-poor populations stronger. Thus one should use the results
shown here for more metal-poor populations carefully. Because it is
impossible to give the formulae for all colours, we give some
formulae for calculating the effects of binary interactions on 12
important colours, which are sensitive to stellar age or
metallicity, according to the work of \cite{Li:2007colourpairs}. As
a result, The fitting formulae for these colours are obtained. The
12 colours are $(B-V)$, $(V-K)$, $(I-H)$, $(R-K)$, $(B-K)$, $(I-K)$,
$(u-r)$, $(r-K)$, $(u-R)$, $(u-K)$, $(z-K)$, and
$(g-J)$\footnote{Colours $(r-K)$, $(u-R)$, $(u-K)$, $(z-K)$, and
$(g-J)$ are composite colours. The $UBVRIJHK$ magnitudes are on
Johnson system, and $ugriz$ magnitudes are on SDSS-$ugriz$ system.}.
Note that $(B-V)$, $(u-r)$, $(u-R)$, and $(z-K)$ are more sensitive
to stellar age and the others to metallicity. Our work shows that
the changes of the above colours caused by binary interactions can
be expressed as
\begin{equation}
    \Delta I' = \sum_{i=1}^{4}
    {{\rm C}_{i}}t^{i-1},
\end{equation}
where $\Delta I'$ is the change of colours caused by binary
interactions, and $t$ is stellar age. The coefficients of the
equation are shown in Table 3. Note that the results for populations
with both Salpeter IMF (standard investigation) and Chabrier IMF are
listed in the table. We can find that equation (2) does not include
the metallicity of populations. The reason is that there is no clear
trend for different metallicities. The fitting of the effects of
binary interactions on 12 colours are shown in Figs. 4, 5, and 6. As
we see, the fitting formulae can give average colour changes caused
by binary interactions. However, because the results calculated
using equation (2) have typical errors about 0.02\,mag, some
additional uncertainties may be brought into the results of stellar
population studies.

\begin{table}[]
\caption[]{Coefficients for equation (2). $UBVRIJHKLMN$ magnitudes
are on Johnson system, and $ugriz$ magnitudes are on SDSS-$ugriz$
system.} \label{Tab:4}
\begin{center}\begin{tabular}{c|cccc|cccc}
\hline\hline\noalign{\smallskip}%\scriptsize
IMF    &            &Salpeter     &           &           &            &Chabrier     &           &\\
\hline \multicolumn{9}{c}{Age $<$ 4.2\,Gyr}\\
\hline
Colour &${\rm C_{1}}$ &${\rm C_{2}}$ &${\rm C_{3}}$ &${\rm C_{4}}$ &${\rm C_{1}}$ &${\rm C_{2}}$ &${\rm C_{3}}$ &${\rm C_{4}}$\\
\hline
(B-V)   &-0.014222   &-0.032764    &0.009111   &-0.000722   &-0.020059   &-0.021895    &0.005569   &-0.000449\\
(V-K)   &-0.080134   &-0.093961    &0.038424   &-0.004241   &-0.086556   &-0.062010    &0.021278   &-0.001931\\
(I-H)   &-0.047703   &-0.049909    &0.021701   &-0.002461   &-0.049641   &-0.033034    &0.011729   &-0.001032\\
(R-K)   &-0.043288   &-0.037439    &0.016264   &-0.001812   &-0.042418   &-0.025391    &0.008674   &-0.000701\\
(B-K)   &-0.094328   &-0.126572    &0.047424   &-0.004947   &-0.106087   &-0.084725    &0.027181   &-0.002419\\
(I-K)   &-0.054306   &-0.055956    &0.024500   &-0.002787   &-0.055634   &-0.036844    &0.012924   &-0.001109\\
(u-r)   &-0.056601   &-0.042940    &0.012347   &-0.001002   &-0.063465   &-0.023075    &0.005139   &-0.000410\\
(r-K)   &-0.072748   &-0.083108    &0.035437   &-0.004014   &-0.077569   &-0.054339    &0.019333   &-0.001782\\
(u-R)   &-0.059219   &-0.047988    &0.014264   &-0.001213   &-0.066913   &-0.026108    &0.006208   &-0.000525\\
(u-K)   &-0.129146   &-0.126222    &0.047806   &-0.005016   &-0.140906   &-0.077425    &0.024514   &-0.002201\\
(z-K)   &-0.043288   &-0.037439    &0.016264   &-0.001812   &-0.042418   &-0.025391    &0.008674   &-0.000701\\
(g-J)   &-0.065198   &-0.092019    &0.035014   &-0.003704   &-0.074778   &-0.061406    &0.020278   &-0.001863\\
\hline\noalign{\bigskip}
\hline \multicolumn{9}{c}{Age $\ge$ 4.2\,Gyr}\\
\hline
Colour &${\rm C_{1}}$ &${\rm C_{2}}$ &${\rm C_{3}}$ &${\rm C_{4}}$ &${\rm C_{1}}$ &${\rm C_{2}}$ &${\rm C_{3}}$ &${\rm C_{4}}$\\
\hline
(B-V)   &-0.069795    &0.008858   &-0.000767    &0.000025   &-0.062030    &0.005088   &-0.000384    &0.000014\\
(V-K)   &-0.157537    &0.014833   &-0.001013    &0.000025   &-0.191178    &0.028283   &-0.002893    &0.000103\\
(I-H)   &-0.087919    &0.010592   &-0.000813    &0.000020   &-0.113589    &0.020964   &-0.002196    &0.000076\\
(R-K)   &-0.078362    &0.010635   &-0.000850    &0.000022   &-0.099112    &0.019317   &-0.002018    &0.000069\\
(B-K)   &-0.222114    &0.020910   &-0.001424    &0.000037   &-0.245483    &0.029974   &-0.002870    &0.000102\\
(I-K)   &-0.100169    &0.012567   &-0.000978    &0.000024   &-0.130611    &0.024976   &-0.002622    &0.000090\\
(u-r)   &-0.119987    &0.009885   &-0.001017    &0.000042   &-0.098610    &0.000563   &-0.000163    &0.000019\\
(r-K)   &-0.135641    &0.013834   &-0.001022    &0.000026   &-0.169588    &0.027336   &-0.002877    &0.000102\\
(u-R)   &-0.126930    &0.010203   &-0.001026    &0.000042   &-0.106134    &0.001228   &-0.000236    &0.000022\\
(u-K)   &-0.255286    &0.021748   &-0.001696    &0.000052   &-0.264095    &0.024796   &-0.002561    &0.000101\\
(z-K)   &-0.078362    &0.010635   &-0.000850    &0.000022   &-0.099112    &0.019317   &-0.002018    &0.000069\\
(g-J)   &-0.154783    &0.014291   &-0.000930    &0.000024   &-0.171114    &0.020427   &-0.001897    &0.000067\\
\noalign{\smallskip}\hline
\end{tabular}\end{center}
\end{table}

%
% one-column-wide figure(occupies half-width of a page)
%  -- This is an old way of graphics inclusion with psfig.sty
%------------------------------------------------------------ Fig1: lightcurve
\begin{figure}
   \vspace{2mm}
   \begin{center}
   \hspace{3mm}\psfig{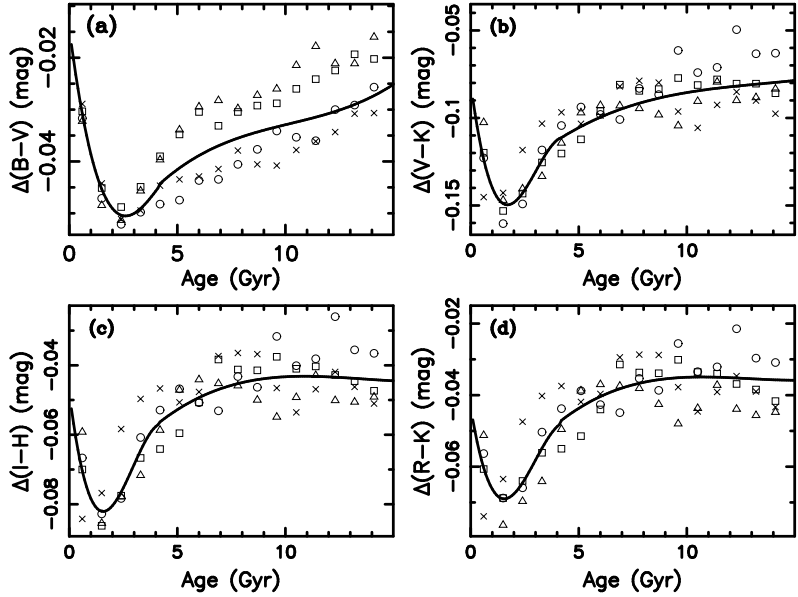}
   \parbox{180mm}{{\vspace{2mm} }}
   \caption{Fitting for the effects of binary interactions on four colours of populations.
   Circles, crosses, squares, and triangles are for the values obtained directly from comparing
   the colours of bs-SSPs and ss-SSPs, for metallicities of 0.004, 0.01, 0.02, and 0.03, respectively.
   Solid lines show the fittings. The y-axis is obtained by subtracting
   the colour of a bs-SSP from that of an ss-SSP (with the same age and metallicity).
   The four panels are for $(B-V)$, $(V-K)$, $(I-H)$, and $(R-K)$, respectively.}
   \label{Fig:lightcurve-ADAri}
   \end{center}
\end{figure}
%

%
% one-column-wide figure(occupies half-width of a page)
%  -- This is an old way of graphics inclusion with psfig.sty
%------------------------------------------------------------ Fig1: lightcurve
\begin{figure}
   \vspace{2mm}
   \begin{center}
   \hspace{3mm}\psfig{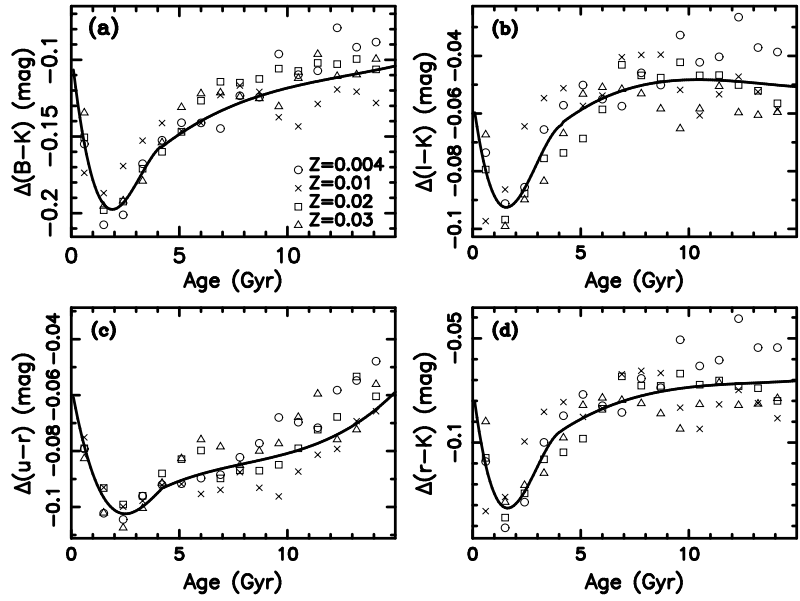}
   \parbox{180mm}{{\vspace{2mm} }}
   \caption{Similar to Fig. 4, but for $(B-K)$, $(I-K)$, $(u-r)$, and $(r-K)$. }
   \label{Fig:lightcurve-ADAri}
   \end{center}
\end{figure}
%

%
% one-column-wide figure(occupies half-width of a page)
%  -- This is an old way of graphics inclusion with psfig.sty
%------------------------------------------------------------ Fig1: lightcurve
\begin{figure}
   \vspace{2mm}
   \begin{center}
   \hspace{3mm}\psfig{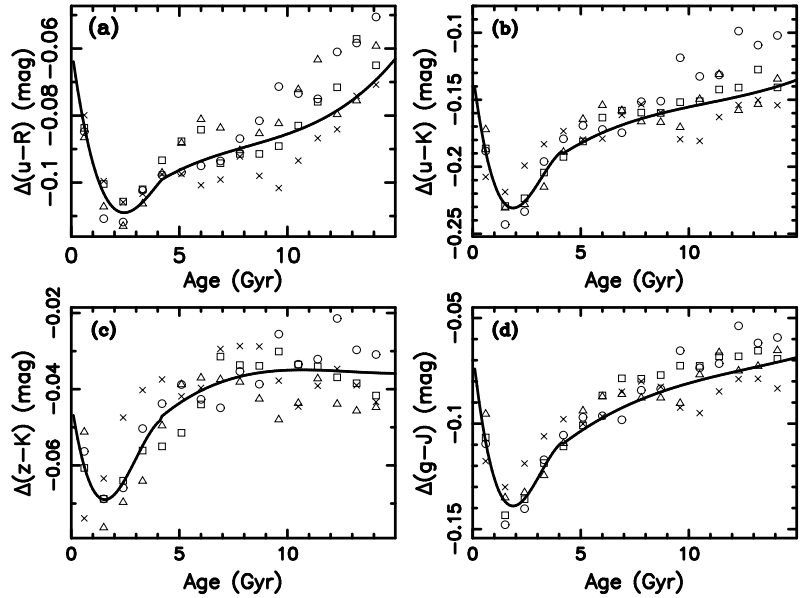}
   \parbox{180mm}{{\vspace{2mm} }}
   \caption{Similar to Fig. 4, but for $(u-R)$, $(u-K)$, $(z-K)$, and $(g-J)$. }
   \label{Fig:lightcurve-ADAri}
   \end{center}
\end{figure}

\section{Discussion and conclusions}
We present some formulae for conveniently computing the changes in
25 Lick indices and 12 colours that are caused by binary
interactions, comparing to the results of single star-stellar
populations (ss-SSPs). It is shown that the fitting formulae
presented in the paper can calculate the changes in Lick indices
caused by binary interactions with small errors and can estimate
similar changes in colours. It is also found that binary
interactions make age-sensitive Lick indices (not only H$\beta$, but
also $H\delta_{\rm A}$, $H\delta_{\rm F}$, $H\gamma_{\rm A}$,
$H\gamma_{\rm F}$) less, while metallicity-sensitive indices larger
compared to those of ss-SSPs. This is useful to estimate the effects
of binary evolution on the results of stellar population studies and
to add the effects of binary interactions into ss-SSP models.
Therefore, when an age-sensitive Lick index is used together with a
metallicity-sensitive index to determine the ages and metallicities
of populations, we will obtain less ages, especially for metal-poor
populations, as the results of \cite{Li:2007binaryeffects}. However,
only binary star-stellar populations (bs-SSPs) and ss-SSPs with four
metallicities ($Z$ = 0.004, 0.01, 0.02, and 0.03) are used in the
work. It makes the results more suitable for studying metal-rich ($Z
\geq$ 0.004) populations, because the differences between integrated
peculiarities of populations with various metallicities seem larger
for metal-poor populations. In addition, although different formulae
are presented for populations with various initial mass functions
(IMFs), the changes calculated via two kinds of formulae (the
formulae for populations with Salpeter and Chabrier IMFs) are
similar for the same population. Thus the changes calculated by the
formulae obtained using populations with Salpeter IMF or Chabrier
IMF can give us some pictures for the effects of binary
interactions. Furthermore, because the Monte Carlo technique used to
generate the binary sample of stellar populations make the evolution
of integrated peculiarities of populations unsmooth, some results,
especially, those for colours, may be somewhat rough. The additional
uncertainties involved should be taken into account. If possible, we
will give more detailed studies in the future.

\begin{acknowledgements}
We thank Profs. Gang Zhao, Xu Zhou, Licai Deng, Xu Kong, Tinggui
Wang, and Li Zhang for useful discussions. This work is supported by
the Chinese National Science Foundation (Grant Nos. 10433030,
10521001, 2007CB815406).
\end{acknowledgements}

%\bibliography{tex}
%\bibliographystyle{apj}

\label{lastpage}

%\end{document}
%%==^..^============== the END of cjaa.tex ===================^_^==

\newpage
\appendix
\section{Coefficients for calculating the effects of binary interactions on 25 Lick indices of populations with Chabrier IMF}

\begin{table}[]
\caption[]{Coefficients for equation (1). The coefficients are
obtained via stellar populations with Chabrier IMF and can be used
for populations younger than 3.5\,Gyr (Age $<$ 3.5\,Gyr).}
\label{Tab:4}
\begin{center}\begin{tabular}{lcrrrrr}
\hline\hline\noalign{\smallskip}%\scriptsize
Index&$j$ &${\rm C_{1j}}$ &${\rm C_{2j}}$ &${\rm C_{3j}}$ &${\rm C_{4j}}$ &${\rm C_{5j}}$\\
\hline
                  &1 &    0.0022919 &    0.0011051 &   -0.0063997 &    0.0015981 &   -0.0001080 \\
CN$_{\rm 1}$      &2 &    0.6329579 &   -0.9595824 &    0.2427491 &   -0.0011151 &   -0.0025634 \\
                  &3 &   -4.5940253 &    7.9284009 &   -3.9250539 &    0.6480545 &   -0.0256989 \\
\hline                                                                                          \\
                  &1 &    0.0012774 &    0.0032128 &   -0.0069053 &    0.0016921 &   -0.0001156 \\
CN$_{\rm 2}$      &2 &    0.6334017 &   -1.0656913 &    0.3400438 &   -0.0309074 &   -0.0000476 \\
                  &3 &   -4.4675288 &    9.5105511 &   -5.7521116 &    1.2233685 &   -0.0747440 \\
\hline                                                                                          \\
                  &1 &    0.0795038 &   -0.1571070 &    0.0740613 &   -0.0128663 &    0.0007442 \\
Ca4227            &2 &  -11.9558663 &   20.8851588 &   -9.0611781 &    1.3421325 &   -0.0622930 \\
                  &3 &  342.8235941 & -529.6545151 &  161.4994646 &  -12.9148165 &   -0.1134096 \\
\hline                                                                                          \\
                  &1 &    0.7660019 &   -1.5947618 &    0.6186025 &   -0.0971560 &    0.0052551 \\
G4300             &2 &  -87.0440009 &  155.4728799 &  -65.2201097 &    9.5672170 &   -0.4355927 \\
                  &3 & 2358.1318663 &-4145.5253913 & 1656.4267226 & -221.2634252 &    8.6149062 \\
\hline                                                                                          \\
                  &1 &    0.6089788 &   -0.9690479 &    0.3009590 &   -0.0329757 &    0.0009352 \\
Fe4383            &2 &  -72.6995453 &  111.7196061 &  -45.0101690 &    6.0333694 &   -0.2331824 \\
                  &3 & 2100.1581312 &-3361.5113576 & 1335.7159550 & -174.8630448 &    6.5371492 \\
\hline                                                                                          \\
                  &1 &    0.0681495 &   -0.1369378 &    0.0592712 &   -0.0105410 &    0.0006446 \\
Ca4455            &2 &   -7.3288591 &   13.2396365 &   -7.0194851 &    1.2883776 &   -0.0767611 \\
                  &3 &  258.9163092 & -510.1718798 &  260.9018384 &  -46.2914894 &    2.6884347 \\
\hline                                                                                          \\
                  &1 &    0.2564424 &   -0.5328445 &    0.2229634 &   -0.0364862 &    0.0020371 \\
Fe4531            &2 &  -26.0845865 &   42.1057901 &  -17.3504899 &    2.4029055 &   -0.0987445 \\
                  &3 &  851.0453910 &-1492.6215470 &  616.0271336 &  -86.7449029 &    3.7433351 \\
\hline                                                                                          \\
                  &1 &    0.1275239 &   -0.2791766 &    0.1024538 &   -0.0161675 &    0.0009372 \\
Fe4668            &2 &   -5.2389930 &    7.6979073 &   -2.3867283 &    0.1846930 &   -0.0012464 \\
                  &3 &  169.0223930 & -536.0764913 &  339.6448905 &  -65.7521400 &    4.1760430 \\
\hline                                                                                          \\
                  &1 &   -0.4293708 &    0.8954661 &   -0.3155113 &    0.0432326 &   -0.0020233 \\
H$_\beta$         &2 &   30.1096288 &  -49.0583758 &   16.8580314 &   -1.8838154 &    0.0479103 \\
                  &3 & -767.8393985 & 1229.9316335 & -430.0651511 &   45.5750506 &   -0.8795314 \\
\hline                                                                                          \\
                  &1 &    0.2395560 &   -0.3782478 &    0.1010824 &   -0.0116568 &    0.0005056 \\
Fe5015            &2 &   -3.3348359 &  -34.6665677 &   26.0424965 &   -5.8877139 &    0.4132806 \\
                  &3 &  413.3734390 &   45.5824955 & -217.1527754 &   66.7394428 &   -5.4107732 \\
\hline                                                                                          \\
                  &1 &    0.0057639 &   -0.0122970 &    0.0048864 &   -0.0007787 &    0.0000435 \\
Mg$_{\rm 1}$      &2 &   -0.3355850 &    0.6283524 &   -0.2850889 &    0.0319693 &   -0.0005680 \\
                  &3 &   13.7239823 &  -27.6095038 &   10.7666288 &   -1.0488923 &    0.0060013 \\
\hline                                                                                          \\
                  &1 &    0.0123326 &   -0.0261803 &    0.0111134 &   -0.0018290 &    0.0001038 \\
Mg$_{\rm 2}$      &2 &   -0.7115473 &    1.0153443 &   -0.3612812 &    0.0205355 &    0.0016030 \\
                  &3 &   27.2273352 &  -44.1981147 &   13.4974485 &   -0.5084821 &   -0.0847616 \\
\hline                                                                                          \\
                  &1 &    0.1259541 &   -0.2256908 &    0.0880777 &   -0.0124308 &    0.0005828 \\
Mg$_{\rm b}$      &2 &   -8.2670312 &    5.7969056 &    0.6635323 &   -0.9033180 &    0.1008251 \\
                  &3 &  311.7946049 & -348.4131106 &   36.5555001 &   20.8760850 &   -2.8507846 \\
\hline                                                                                          \\
                  &1 &    0.1526957 &   -0.3633561 &    0.1689542 &   -0.0304730 &    0.0018686 \\
Fe5270            &2 &  -16.8516642 &   33.9221551 &  -18.3927017 &    3.5096217 &   -0.2189850 \\
                  &3 &  587.4969012 &-1243.1886864 &  678.5522252 & -131.0104877 &    8.3026556 \\
\noalign{\smallskip}

\noalign{\smallskip}\hline
\end{tabular}\end{center}
\end{table}

\addtocounter{table}{-1}
\begin{table}[]
\centering \caption[]{--continued.} \label{Tab:4}
\begin{center}\begin{tabular}{lcrrrrr}
\hline\hline\noalign{\smallskip}%\scriptsize
Index&$j$ &${\rm C_{1j}}$ &${\rm C_{2j}}$ &${\rm C_{3j}}$ &${\rm C_{4j}}$ &${\rm C_{5j}}$\\
\hline                                                                                        \\
                  &1 &    0.2076133 &   -0.4791791 &    0.2222679 &   -0.0377229 &    0.0021281 \\
Fe5335            &2 &  -32.4556974 &   65.1973802 &  -31.8891247 &    5.4057781 &   -0.2980594 \\
                  &3 &  935.2591701 &-1881.6098576 &  872.5249261 & -142.2695305 &    7.5855211 \\
\hline                                                                                          \\
                  &1 &    0.0953714 &   -0.2161430 &    0.0970755 &   -0.0171741 &    0.0010419 \\
Fe5406            &2 &   -3.9285567 &    6.4393293 &   -3.6907976 &    0.6355834 &   -0.0343946 \\
                  &3 &  174.8445638 & -345.4890330 &  172.7005100 &  -28.2035654 &    1.4789489 \\
\hline                                                                                          \\
                  &1 &    0.0665533 &   -0.1333817 &    0.0460766 &   -0.0056987 &    0.0002133 \\
Fe5709            &2 &   -8.4184254 &   11.6155577 &   -2.8870447 &    0.0168183 &    0.0277514 \\
                  &3 &  229.4380018 & -346.7249858 &  109.9625205 &   -7.5674339 &   -0.2639790 \\
\hline                                                                                          \\
                  &1 &    0.0724796 &   -0.1553226 &    0.0682675 &   -0.0111777 &    0.0006131 \\
Fe5782            &2 &  -12.3652716 &   23.1281644 &  -10.7411540 &    1.7511767 &   -0.0935373 \\
                  &3 &  379.6596477 & -665.9564600 &  268.9228630 &  -38.4448161 &    1.7703055 \\
\hline                                                                                          \\
                  &1 &    0.0714688 &   -0.1651977 &    0.0664805 &   -0.0103341 &    0.0005679 \\
Na$_{\rm D}$      &2 &   -1.5605878 &   -4.9385375 &    5.3394247 &   -1.5674185 &    0.1304109 \\
                  &3 &  148.3896060 &  -24.0899505 & -130.0698652 &   49.4638805 &   -4.4865514 \\
\hline                                                                                          \\
                  &1 &   -0.0115706 &    0.0112891 &   -0.0010864 &   -0.0006067 &    0.0000840 \\
TiO$_{\rm 1}$     &2 &    2.0705973 &   -2.9503555 &    0.8324854 &   -0.0401368 &   -0.0040348 \\
                  &3 &  -44.5370053 &   67.5680594 &  -20.7546877 &    1.4636658 &    0.0462995 \\
\hline                                                                                          \\
                  &1 &   -0.0125496 &    0.0097243 &   -0.0001968 &   -0.0008541 &    0.0001063 \\
TiO$_{\rm 2}$     &2 &    2.4479466 &   -3.6429429 &    1.1610628 &   -0.0929928 &   -0.0014161 \\
                  &3 &  -49.2619296 &   79.6645168 &  -27.3299096 &    2.6274999 &   -0.0183345 \\
\hline                                                                                          \\
                  &1 &   -0.1701616 &    0.1253346 &    0.2496936 &   -0.0688490 &    0.0048300 \\
H$\delta_{\rm A}$ &2 &  -22.7201592 &   34.6318625 &  -16.9398410 &    2.8138566 &   -0.1414102 \\
                  &3 &   35.0832106 & -330.3012402 &  581.7572135 & -167.8682957 &   12.8850332 \\
\hline                                                                                          \\
                  &1 &   -1.5856131 &    2.7085039 &   -0.9113125 &    0.1227389 &   -0.0055250 \\
H$\gamma_{\rm A}$ &2 &  180.2526812 & -282.6477080 &  118.3329534 &  -17.2376118 &    0.7715612 \\
                  &3 &-5010.4626342 & 7614.1515475 &-2881.3416065 &  351.0737254 &  -11.0363129 \\
\hline                                                                                          \\
                  &1 &   -0.1251870 &    0.1470139 &    0.1086295 &   -0.0343830 &    0.0025515 \\
H$\delta_{\rm F}$ &2 &  -17.3098016 &   30.7053467 &  -17.0765539 &    3.2519508 &   -0.1978084 \\
                  &3 &  137.3583135 & -435.8552796 &  470.4948769 & -124.2942197 &    9.3212486 \\
\hline                                                                                          \\
                  &1 &   -0.6254725 &    1.1213832 &   -0.3612095 &    0.0484349 &   -0.0022475 \\
H$\gamma_{\rm F}$ &2 &   53.4441919 &  -86.7513473 &   36.9363160 &   -5.4971981 &    0.2510974 \\
                  &3 &-1532.4775410 & 2361.4227142 & -884.2807841 &  102.2525869 &   -2.6516141 \\
\noalign{\smallskip}\hline
\end{tabular}\end{center}
\end{table}

\begin{table}[]
\caption[]{Similar to Table A.1, but for stellar populations older
than 3.5\,Gyr (Age $\geq$ 3.5\,Gyr).} \label{Tab:4}
\begin{center}\begin{tabular}{lcrrrrr}
\hline\hline\noalign{\smallskip}%\scriptsize
Index&$j$ &${\rm C_{1j}}$ &${\rm C_{2j}}$ &${\rm C_{3j}}$ &${\rm C_{4j}}$ &${\rm C_{5j}}$\\
\hline
                  &1 &    0.0050070 &   -0.0101123 &    0.0016866 &   -0.0001101 &    0.0000026 \\
CN$_{\rm 1}$      &2 &   -0.7387795 &   -0.0268996 &    0.0467357 &   -0.0067710 &    0.0002559 \\
                  &3 &    1.8158227 &    8.1027523 &   -2.3010584 &    0.2598589 &   -0.0092777 \\
\hline                                                                                          \\
                  &1 &    0.0048267 &   -0.0086639 &    0.0014493 &   -0.0000926 &    0.0000021 \\
CN$_{\rm 2}$      &2 &   -0.6713808 &   -0.0767864 &    0.0454875 &   -0.0063150 &    0.0002435 \\
                  &3 &    2.4347829 &    7.4817292 &   -1.9820797 &    0.2264804 &   -0.0083295 \\
\hline                                                                                          \\
                  &1 &    0.0253654 &   -0.0209602 &    0.0048307 &   -0.0004936 &    0.0000168 \\
Ca4227            &2 &   -3.5901788 &    1.5319737 &   -0.4799447 &    0.0598457 &   -0.0022187 \\
                  &3 &  108.6370369 & -107.2440863 &   25.4608949 &   -2.5477372 &    0.0859891 \\
\hline                                                                                          \\
                  &1 &    0.2092102 &   -0.4271461 &    0.0709524 &   -0.0053369 &    0.0001564 \\
G4300             &2 &  -32.0561313 &   13.8970486 &   -0.9523866 &    0.0227085 &   -0.0010081 \\
                  &3 &  388.0491119 & -107.1700035 &   -9.9303420 &    2.4271319 &   -0.0776931 \\
\hline                                                                                          \\
                  &1 &    0.0852039 &   -0.2016152 &    0.0323234 &   -0.0021218 &    0.0000535 \\
Fe4383            &2 &  -12.6431474 &   -2.2748662 &    1.1500142 &   -0.1342407 &    0.0045588 \\
                  &3 &   60.8341421 &  179.0155270 &  -49.0631558 &    5.2647148 &   -0.1806644 \\
\hline                                                                                          \\
                  &1 &    0.0439274 &   -0.0430992 &    0.0064975 &   -0.0004070 &    0.0000095 \\
Ca4455            &2 &   -6.6685854 &    1.4457768 &   -0.0535448 &   -0.0104817 &    0.0006266 \\
                  &3 &  114.6093932 &   -5.2478791 &   -4.3355245 &    0.8004669 &   -0.0341767 \\
\hline                                                                                          \\
                  &1 &    0.0586416 &   -0.1084756 &    0.0185745 &   -0.0013543 &    0.0000372 \\
Fe4531            &2 &  -16.5223730 &    5.9250743 &   -0.6822331 &    0.0345448 &   -0.0007249 \\
                  &3 &  327.8776772 & -131.2118033 &   13.2546833 &   -0.3619352 &   -0.0008992 \\
\hline                                                                                          \\
                  &1 &    0.0579868 &   -0.0883225 &    0.0125230 &   -0.0005465 &    0.0000058 \\
Fe4668            &2 &  -14.2549909 &    3.6173449 &   -0.0066150 &   -0.0775483 &    0.0041877 \\
                  &3 &  256.6237954 &   70.1814681 &  -29.9674944 &    4.6116691 &   -0.1954322 \\
\hline                                                                                          \\
                  &1 &    0.1188078 &    0.1462153 &   -0.0301692 &    0.0023434 &   -0.0000653 \\
H$_\beta$         &2 &    5.2527398 &   -5.1845420 &    0.5595370 &   -0.0222041 &    0.0004090 \\
                  &3 &   -6.8462900 &   -4.4966224 &    9.4442175 &   -1.3358200 &    0.0457839 \\
\hline                                                                                          \\
                  &1 &    0.0327476 &   -0.1444852 &    0.0252383 &   -0.0018120 &    0.0000483 \\
Fe5015            &2 &  -22.2627396 &    7.4518761 &   -0.7875696 &    0.0244834 &    0.0000504 \\
                  &3 &  369.7815953 &  -72.0823837 &   -1.2940653 &    1.1387862 &   -0.0545511 \\
\hline                                                                                          \\
                  &1 &   -0.0029387 &   -0.0005618 &    0.0000571 &   -0.0000025 &    0.0000001 \\
Mg$_{\rm 1}$      &2 &    0.0681456 &   -0.2239214 &    0.0461259 &   -0.0037007 &    0.0001017 \\
                  &3 &   -1.7505218 &    4.5445049 &   -1.1536134 &    0.1112986 &   -0.0034731 \\
\hline                                                                                          \\
                  &1 &   -0.0022995 &   -0.0025539 &    0.0005160 &   -0.0000415 &    0.0000013 \\
Mg$_{\rm 2}$      &2 &   -0.3140984 &   -0.1102010 &    0.0232351 &   -0.0016550 &    0.0000414 \\
                  &3 &    3.8491416 &    2.2473119 &   -0.6944256 &    0.0739731 &   -0.0024430 \\
\hline                                                                                          \\
                  &1 &    0.0165715 &   -0.0215425 &    0.0043674 &   -0.0002914 &    0.0000068 \\
Mg$_{\rm b}$      &2 &   -8.0187674 &    0.1612416 &    0.0564198 &   -0.0091935 &    0.0004596 \\
                  &3 &  158.6306124 &  -15.8632521 &    0.7547564 &    0.1990004 &   -0.0142351 \\
\hline                                                                                          \\
                  &1 &   -0.0053182 &   -0.0449585 &    0.0074887 &   -0.0004761 &    0.0000116 \\
Fe5270            &2 &   -6.2888460 &    0.8351887 &    0.0980695 &   -0.0229461 &    0.0009453 \\
                  &3 &  106.4680655 &   11.8581491 &   -9.0607405 &    1.2154258 &   -0.0455243 \\
\noalign{\smallskip}

\noalign{\smallskip}\hline
\end{tabular}\end{center}
\end{table}

\addtocounter{table}{-1}
\begin{table}[]
\centering \caption[]{--continued.} \label{Tab:4}
\begin{center}\begin{tabular}{lcrrrrr}
\hline\hline\noalign{\smallskip}%\scriptsize
Index&$j$ &${\rm C_{1j}}$ &${\rm C_{2j}}$ &${\rm C_{3j}}$ &${\rm C_{4j}}$ &${\rm C_{5j}}$\\
                  &1 &   -0.0299036 &   -0.0168634 &    0.0034824 &   -0.0003504 &    0.0000128 \\
Fe5335            &2 &   -0.8764299 &   -2.3631112 &    0.3216860 &   -0.0034947 &   -0.0004859 \\
                  &3 &  -19.1581245 &   20.3079448 &    1.5735482 &   -0.7362636 &    0.0365434 \\
\hline                                                                                          \\
                  &1 &   -0.0273466 &   -0.0191557 &    0.0032597 &   -0.0002321 &    0.0000069 \\
Fe5406            &2 &   -1.6859288 &   -1.4089294 &    0.3578634 &   -0.0309966 &    0.0008859 \\
                  &3 &    3.2750120 &   49.7275784 &  -12.4265243 &    1.1493382 &   -0.0351427 \\
\hline                                                                                          \\
                  &1 &    0.0058300 &   -0.0239914 &    0.0038273 &   -0.0002377 &    0.0000055 \\
Fe5709            &2 &   -3.1695351 &    1.3275974 &   -0.1282929 &    0.0015248 &    0.0001363 \\
                  &3 &   62.3172407 &   -5.6986762 &   -1.2147039 &    0.3086961 &   -0.0142722 \\
\hline                                                                                          \\
                  &1 &   -0.0351759 &    0.0072524 &   -0.0009035 &   -0.0000295 &    0.0000040 \\
Fe5782            &2 &    2.9123930 &   -2.1632509 &    0.2392269 &    0.0009764 &   -0.0005652 \\
                  &3 &  -87.2972968 &   23.4730212 &    0.8426133 &   -0.6095760 &    0.0321487 \\
\hline                                                                                          \\
                  &1 &   -0.0349510 &   -0.0109978 &    0.0032700 &   -0.0003007 &    0.0000101 \\
Na$_{\rm D}$      &2 &   -3.1483417 &   -0.0799733 &    0.0306158 &    0.0017984 &   -0.0001874 \\
                  &3 &   74.8532596 &  -28.0695426 &    1.8470812 &    0.0045464 &   -0.0013129 \\
\hline                                                                                          \\
                  &1 &   -0.0019580 &   -0.0001078 &    0.0001071 &   -0.0000133 &    0.0000005 \\
TiO$_{\rm 1}$     &2 &   -0.0663381 &    0.0341998 &   -0.0108762 &    0.0008305 &   -0.0000198 \\
                  &3 &    3.2582308 &   -1.5575893 &    0.2498990 &   -0.0085626 &   -0.0000514 \\
\hline                                                                                          \\
                  &1 &   -0.0055418 &    0.0002662 &    0.0000839 &   -0.0000138 &    0.0000006 \\
TiO$_{\rm 2}$     &2 &    0.1861284 &   -0.0836849 &    0.0091382 &   -0.0007762 &    0.0000260 \\
                  &3 &   -4.2466215 &    3.0328182 &   -0.7327127 &    0.0800970 &   -0.0027508 \\
\hline                                                                                          \\
                  &1 &   -0.4834854 &    0.6747627 &   -0.1136504 &    0.0081802 &   -0.0002208 \\
H$\delta_{\rm A}$ &2 &   48.2162849 &   -8.6259083 &   -0.5680772 &    0.1330894 &   -0.0044290 \\
                  &3 & -391.8235480 &  -44.3395429 &   41.4350353 &   -5.4239881 &    0.1828454 \\
\hline                                                                                          \\
                  &1 &   -0.5340026 &    0.8499248 &   -0.1407715 &    0.0102053 &   -0.0002837 \\
H$\gamma_{\rm A}$ &2 &   57.5218630 &  -12.4432398 &   -1.0323349 &    0.2184516 &   -0.0069865 \\
                  &3 & -381.7703663 & -237.3724815 &  109.5206537 &  -12.9300524 &    0.4258356 \\
\hline                                                                                          \\
                  &1 &   -0.2328216 &    0.3790464 &   -0.0645870 &    0.0047630 &   -0.0001318 \\
H$\delta_{\rm F}$ &2 &   28.4618861 &   -8.4343118 &    0.2882612 &    0.0224536 &   -0.0007588 \\
                  &3 & -256.8427724 &   41.7473144 &   12.0094480 &   -2.0067178 &    0.0654486 \\
\hline                                                                                          \\
                  &1 &   -0.2081433 &    0.4185126 &   -0.0709583 &    0.0052155 &   -0.0001461 \\
H$\gamma_{\rm F}$ &2 &   26.6976093 &   -8.4741360 &    0.0074237 &    0.0631139 &   -0.0020772 \\
                  &3 & -176.6172339 &  -82.1781969 &   45.6384532 &   -5.4871418 &    0.1779060 \\
\noalign{\smallskip}

\noalign{\smallskip}\hline
\end{tabular}\end{center}
\end{table}

\end{document}